\documentclass{article}
\usepackage{cite}
\usepackage{pdfpages}
\usepackage{xcolor}
\usepackage{authblk}
\usepackage{graphicx}
\usepackage{amssymb}
\usepackage{amsmath}
\usepackage[top=5cm, bottom=5cm, left=3cm, right=2cm, headsep=1.5cm, heightrounded=true]{geometry}
\usepackage{fancyhdr}
\graphicspath{ {{images/}} }

\title{ 4D-EGB Black Holes in RPS Thermodynamics }

\author{Y. Ladghami\thanks{ \texttt{yahya.ladghami@ump.ac.ma}}}
\author{B. Asfour\thanks{ \texttt{brahim.asfour@ump.ac.ma}}}
\author{A. Bouali\thanks{ \texttt{a1.bouali@ump.ac.ma}}}
\author{A. Errahmani\thanks{ \texttt{ahmederrahmani1@yahoo.fr}}}
\author{T. Ouali\thanks{ \texttt{ouali1962@gmail.com}}}
\affil {Laboratory of Physics of Matter and Radiation, Mohammed I University, BP 717, Oujda, Morocco}
\begin{document}
\maketitle

\begin{abstract}
 In this paper, we study thermodynamics of charged and uncharged 4-Dimension  Einstein-Gauss-Bonnet (4D-EGB) black holes. The context of this study is the Visser's  holographic thermodynamics with a fixed anti-de Sitter radius and a variable Newton constant known as restricted phase space thermodynamics (RPST). Our setup is constructed by using the AdS/CFT correspondence and by introducing  a conjugate quantity of the Gauss-Bonnet parameter. By this ansatz, we conclude that the Gauss-Bonnet action multiplied by a temperature, behaves as a free energy. We derive the conjugate quantities corresponding to the first law in the RPST formalism.  
 The study of the $T-S$ processes and the effect of the Gauss-Bonnet constant, $\alpha$, show that thermodynamic properties of charged black holes depend on the Gauss-Bonnet term and the charge of black holes.  For an uncharged black holes,  the  effect of Gauss-Bonnet becomes crucial, as it behaves as a  charged black hole with an effective charge. Finally, we find that the Hawking-Page phase transition occurs between a large black hole and a thermal AdS space.
\end{abstract}
\section{Introduction}
\par Black holes and their thermodynamics  provide a fertile ground for research, especially after Hawking and Bekenstein's work on the temperature and the entropy of black holes \cite{Hawking:1975vcx,B1}, in addition to the four laws of black hole thermodynamics, through the thermodynamic analogy of the ordinary systems \cite{B72,Bardeen1973,1976black}. This  thermodynamic is known as  traditional black hole thermodynamics (TBHT), for which,  the mass of the black hole is considered as its internal energy. The most important results of TBHT are the Hawking-Page phase transition for the space-time with a negative cosmological constant ($\Lambda<0$) i.e an anti-de Sitter space-time  (AdS) \cite{H-P}.  Another important formalism in the study of black holes is called extended phase  space thermodynamics (EPST) started for the first time in \cite{kastor2009enthalpy}. This formalism is also called "Black hole chemistry" \cite{kubizvnak2015black,kubizvnak2017black}. A number of works have been published in this regard  \cite{Dolan_2011,Cai_2013,Kubiz_k_2012,Dol11,b22,wang2020extended,hendi2013extended,estrada2020thermodynamic}. In the EPST, the mass of black holes is interpreted as an enthalpy and thermodynamic parameters such as the pressure, P, and the volume, V, have been interpenetrated as  thermodynamic variables,  where the pressure is related to the cosmological constant via $P=-\Lambda/ 8\pi G$ \cite{kubizvnak2015black}. AdS/CFT correspondence, speculated by Maldacena  \cite{Maldacena_1999},  has played an important role to understand the behavior of black hole and it is used as a framework to introduce a new thermodynamic approach. In this background, the AdS black hole is equivalent to a thermal state in the conformal field theories (CFT) \cite{Hashimoto_2020}. Recently, within the framework of the AdS/CFT correspondence and based on the Visser’s thermodynamics  \cite{cVisser_2022}, a new thermodynamic called restricted phase space thermodynamics (RPST) was initiated by Gao et al in  \cite{bGao_2022}. In RPST, black hole masses are reinterpreted as internal energy. However, instead of the pressure  and the volume  characterizing the  EPST formalism, a new pair of conjugate thermodynamic variables characterizing the RPST formalism has been introduced. This new  conjugates thermodynamic variables, are the  chemical potential (color susceptibility) $\mu$,  and the central charge C which represents the number degrees of freedom  (number of colors) in CFT \cite{cVisser_2022,Karch_2015}. The role of central charge C is  similar to that of the number of particles in statistical physics of ordinary matter \cite{Kerr22}. The central charge  is related to the Newton's constant G and the AdS curvature radius $l$ via $C=l^{d-2}/G$, where $d$ is the dimension of the AdS space-time \cite{Karch_2015, bGao_2022}. Homogeneity of Smarr relation,  thermodynamic processes and  phase transitions, were studied in the  RN-AdS,  Kerr-AdS, BTZ  and Taub-NUT black holes   in the context of  RPST \cite{bGao_2022,Kerr22,sadeghi2022rps,bai2022revisit,kong2022restricted}.
\\

Weyl \cite{1} and Eddington \cite{2} first suggested modifications of General Relativity (GR) in 1918, and the idea has stuck around ever since. The current concept aims to develop GR alternatives for some cosmological issues, such as the cosmic acceleration in the early and  late universe.
One of these alternative theories is the Gauss-Bonnet gravity \cite{3,4},  in which an arbitrary function $f(\mathcal{G})$ is included in the gravitational action, where $\mathcal{G}$ is the Gauss-Bonnet (GB) term. However, gravitational dynamics in 4-dimensional spacetime are unaffected by the GB term. For this reason, D. Glavan and C. Lin have recently proposed a novel 4-dimensional Einstein-Gauss-Bonnet (4D-EGB) gravity \cite{Glavan_2020}, which is founded on the idea of multiplying the GB term by the factor $1/(d - 4)$ in order to circumvent the strict requirements of Lovelock’s theorem \cite{6} and avoid Ostrogradsky instability. Additionally, in recent years, this new approach has developed into a useful framework for researching physics and thermodynamics of black holes.  For instance, vacuum solution for charged black holes, thermodynamic properties\cite{aFernandes_2020}, extended thermodynamics of black holes \cite{E20,7}, a charged  black hole with entangled particles and antiparticles \cite{bousder2021particle}, stable circular orbit and shadow of the black holes \cite{g3s},  power spectra of the Hawking radiation in de-Sitter spacetime \cite{z14}, stability of black holes \cite{1R} and non-linear charged planar black holes \cite{7M,y0}.

As far as we know, even with the number  of applications of 4D-EGB, there is no study so far about the RPST formalism in this approach. This motivates us to consider the RPST formalism in the 4D-EGB gravity background. A second motivation is to study effects of the Gauss-Bonnet constant on thermodynamic behaviors of a charged and uncharged black holes in both approaches and to complete the  study of black holes in the RPST formalism already made in \cite{bGao_2022,Kerr22,sadeghi2022rps,bai2022revisit,kong2022restricted}.
\\

The paper is organised as follows: In Section \ref{.1}, we present the 4D-EGB black holes solution in AdS space-time. In Section \ref{.2}, we construct an holographic thermodynamics with a fixed AdS radius for 4D-EGB gravity. In section \ref{.3}, the first law and thermodynamic variables are deduced from RPST. In section \ref{.4}, we look at thermodynamic properties of the charged 4D-EGB black hole in the RPS thermodynamics by $T-S$ processes. In section \ref{.5}, we study uncharged 4D-EGB black holes. In section \ref{.7},  we discuss our main conclusions. 
In this paper, we adopt the unit $ \hbar= c = k_B = 1$.

\section{4D-EGB Black Holes in AdS space-time}
\label{.1}
The action of the Gauss-Bonnet  gravity in a d-dimensional space-time with a negative cosmological constant is given by \cite{aFernandes_2020} 
\begin{equation}
	\label{eq1}
I=\frac{1}{16 \pi G} \int d^d x\left(R - 2\Lambda-F_{\mu \nu} F^{\mu \nu} +\frac{\alpha}{d-4} \mathcal{G}\right),
\end{equation}
where the GB term writes  
\begin{equation}
\mathcal{G}=R_{\mu \nu \rho \sigma} R^{\mu \nu \rho \sigma}-4 R_{\mu \nu} R^{\mu \nu}+R^2,
\end{equation}
$\alpha$ is  Gauss-Bonnet coupling with dimension of [length]$^2$ and represents ultraviolet
corrections to the Einstein theory, $G$ is Newton’s constant and the cosmological constant is given by  
\begin{equation}
\Lambda=-\frac{(d-1)(d-2)}{2 l^2}.
\end{equation}
 The AdS radius is given by $l$ and $  F_{\mu \nu}=\partial_\mu A_\nu-\partial_\nu A_\mu  $ is the Maxwell strength field. The black hole metric, by adopting the limit $d\longrightarrow 4$, is given by \cite{Glavan_2020,Fernandes_2022}

\begin{equation}
d s^2=-f(r) d t^2+\frac{1}{f(r)} d r^2+r^2\left(d \theta^2+\sin ^2 \theta d \phi^2\right),
\end{equation}
where the metric function is given by
\begin{equation}
\label{5}
f(r)=1+\frac{r^2}{2 \alpha}\left(1-\sqrt{1+4 \alpha\left(\frac{2 M G}{r^3}-\frac{Q^2 G}{r^4}-\frac{1}{l^2}\right)} \right),
\end{equation}
where Q is the electric charge and M is the black hole's mass. Solving the equation $f(r)=0$, we find two horizons:  the event horizon $r_+$ and the Cauchy horizon $r_-$. Using Eq. (\ref{5})  and the equation for the even horizon,  $f(r_{+}) = 0$, the black hole's mass is given by

\begin{equation}
\label{88}
M=\frac{r_+}{2G} + \frac{\alpha}{2\, r_+ G} + \frac{r_+^3}{2\, l^2 G} + \frac{ Q^2}{2\, r_+ }.
\end{equation}
The Hawking temperature of the 4D-EGB black hole is given by
\begin{equation}
	\label{Haw}
T=\frac{f'(r_+)}{4 \pi}	= \frac{3 r^4 l^{-2}+r^2  -G  Q^2 -\alpha }{4 \pi   r^3+8 \pi \alpha  r},
\end{equation}
and the electrical potential of the event horizon is defined by
\begin{equation}
    \Phi = \frac{Q}{r_+}.
\end{equation}
In the limit $\alpha \longrightarrow 0$, we recover the mass of Reissner–Nordström black hole in AdS space-time (RN-AdS) \cite{bGao_2022}.

\section{Holographic thermodynamics of black holes  }
\label{.2}
In this section, we will construct holographic thermodynamics of  4D-EGB black hole based on AdS/CFT correspondence. In this construction, the function of a partition in the AdS space-time, $Z_{AdS}$, is equal
to the one of the dual boundary theory i.e. $Z_{CFT}$  \cite{AdS,cVisser_2022}
\begin{equation}
    Z_{AdS} = Z_{CFT}. \label{eq.8}
\end{equation}
At finite temperature, the thermal partition function and the free energy $W$ are related by
\begin{equation}
\label{eq.9}
    W= -T\ln{Z_{CFT}},
\end{equation}
where T is the temperature. In the other hand, the grand canonical  ensemble defines the free energy as 
\begin{equation}
	W=\mu C, \label{eq.10}
\end{equation}   
where the central charge C determines the number of degrees of microscopic freedom in CFT and its conjugate, $\mu$, is considered as a chemical potential.

 The  partition function in the AdS space-time is given by the Euclidean action, $I_E$, of the gravitational theory \cite{dPhysRevD.15.2752}
 \begin{equation}
 	\label{e12}
 I_{\mathrm{E}}=-\ln Z_{\mathrm{AdS}}. 
 \end{equation}
In the 4D-EGB gravity, the Euclidean action $I_E$ is given by Eq. \eqref{eq1} and can be decomposed as
\begin{equation}
\label{14}
I_E= I_{EM} + I_{GB},
\end{equation}
where
\begin{equation}
I_{EM}= \frac{1}{16 \pi G} \int d^d x\left(R - 2\Lambda-F_{\mu \nu} F^{\mu \nu}\right),
\end{equation}
and 
\begin{equation}
 I_{GB}=\frac{1}{16 \pi G} \int d^d x\frac{\alpha}{d-4} \mathcal{G},
\end{equation}
are the Einstein-Maxwell (EM) action and the Gauss-Bonnet action, respectively. The EM action  $I_{EM}$ is given by \cite{dPhysRevD.15.2752}
\begin{equation}
T I_{EM}=M-TS_{BH}-\Phi Q -\Omega J , \label{187}
\end{equation}
where $S_{BH}= {\pi r_+^2}/{G}$ is the Bekenstein-Hawking entropy, T is its conjugate quantity, $\Phi$ is the electrical potential of the event horizon , $\Omega $ is the angular velocity  and  $J$ the  angular momentum.  From now on, we will denote $S_{BH}$ by $S$. By analogy to Eq. (\Ref{187}), we set the GB action  $I_{GB}$ as
\begin{equation}
	\label{GB}
 T I_{GB}= - \alpha \mathcal{A},
\end{equation}
 where we have introduced a new quantity $\mathcal{A}$ as the conjugate of the GB constant $\alpha$. The expression of this new quantity will be given later in terms of thermodynamics and AdS/CFT quantities. The free energy $W$ is expressed as  the partition function  Z such that
\begin{equation}
    Z_{CFT} = e^{-\beta W}, 
\end{equation}
where $\beta = 1/T$.\\
 From Eqs. \eqref{eq.8}- \eqref{e12},  Eqs.  \eqref{187} and  \eqref{GB}, The classical gravitational action writes
\begin{equation}
    Z_{AdS} = e^{-\beta\, \mu \, C}= e^{-\beta\left(W_{EM} + W_{GB}\right)},
\end{equation}
where 
\begin{equation}
    W_{EM}= M-TS-\Phi Q-\Omega J ,
\end{equation}
and 
\begin{equation}
    W_{GB} = - \alpha \mathcal{A}.
\end{equation}
are the Einstein-Maxwell and the Gauss-Bonnet free energies, respectively.\\
Finally, the free energy, can be rewritten as  
\begin{equation}
	\label{eq.16}
	\mu C =  M-TS-\Phi Q- \Omega J -\alpha \mathcal{A}.
\end{equation}
\section{Charged Black Holes}
\label{.3}
In this section, we study in the RPST formalism a charged 4D-EGB black hole i.e. $Q\ne0$ and $J=0$. As we indicated in the introduction, the pressure P and the volume V lose their meanings in the RPS thermodynamics standpoint by introducing the quantity $C$ and its conjugate parameter $\mu$ \cite{bGao_2022,cVisser_2022}.  The relationship between the central charge, the AdS radius  and the Newton’s constant are given by \cite{bGao_2022,Kerr22}
\begin{equation}
C=\frac{l^2}{G}.
\end{equation}
The mass of black holes in  RPST is interpreted as an internal energy \cite{bGao_2022, cVisser_2022}, and from Eq. (\ref{eq.16}), we find the Smarr relation of a charged 4D-EGB black hole  in the RPS formalism as

\begin{equation}
	\label{P7}
 M = T  S + \tilde{\Phi} \tilde{Q} + \mu  C + \mathcal{A}  \alpha,
\end{equation}
where the re-scaled electric potential, $\tilde{\Phi}$,  and the re-scaled  electric charge,  $\tilde{Q}$, are defined by the dual CFT quantities as \cite{bGao_2022,Cong_2021} 
 \begin{equation}
\tilde{Q}=\frac{Q l}{\sqrt{G}},
\end{equation}
and
\begin{equation}
\tilde{\Phi}=\frac{\Phi \sqrt{G}}{l} = \frac{Q \sqrt{G}}{r_+ l},
\end{equation}
respectively. From Smarr's relation, we can say that the mass of a black hole is a function of extensive variables
\begin{equation}
	\label{25}
	M=M(S,\tilde{Q},C,\alpha).
\end{equation}
By differentiating Eq. (\ref{25}), we can formulate the first law in the RPS thermodynamics for a charged 4D-EGB black hole as
\begin{equation}
	\label{10}
	dM = TdS + \Phi d\tilde{Q}  + \mu dC + \mathcal{A} d \alpha	,
	\end{equation}
with the corresponding conjugate quantities 
\begin{equation}
	\label{p}
T=\left(\frac{\partial M}{\partial S}\right)_{\tilde{Q}, C, \alpha}, \qquad \tilde{\Phi}=\left(\frac{\partial M}{\partial \tilde{Q}}\right)_{S, C, \alpha}, \qquad 	\mu=\left(\frac{\partial M}{\partial C}\right)_{S, \tilde{Q}, \alpha}, \quad \text{and} \quad \mathcal{A}=\left(\frac{\partial M}{\partial \alpha}\right)_{S, \tilde{Q}, C}.
\end{equation}
We can also infer the Gibbs-Duheim relation 
\begin{equation}
	-S d T = \tilde{Q} d \Phi   + C d\mu  +  \alpha	d\mathcal{A}.
\end{equation}
 Furthermore, the event horizon radius $r_+$ in terms of the Bekenstein-Hawking entropy and the AdS/CFT quantities writes
\begin{equation}
\label{18}
r_{+}=l\sqrt{\frac{S}{\pi C}}.
\end{equation}
In order to compute  thermodynamic variables in RPST, we rewrite the mass of the 4D-EGB black hole, Eq. (\ref{88}), as
\begin{equation}
	\label{Y}
    M=\frac{S^2+\pi S C+\pi^2 \tilde{Q}^2 + \pi^2 \alpha\, C^2 l^{-2}}{2 \pi^{3 / 2} l \sqrt{S C}}.
\end{equation}
For $\alpha=0$, we recover the mass of the RN-AdS black holes in the RPS formalism \cite{bGao_2022}. The RPST  temperature of the 4D-EGB black hole is defined as a partial derivative of Eq.  (\ref{Y}), i.e
\begin{equation}
\label{16}
T =  \frac{3 S^2+\pi S C-\pi^2 \tilde{Q}^2 - \alpha C^2 \pi^2 l^{-2}}{4 \pi^{3 / 2} l S\sqrt{S C}}.
\end{equation}
This temperature corresponds to the Hawking temperature, Eq. \eqref{Haw}, for $A\gg 8 \pi \alpha$, where $A$ is the area of the event horizon 
.The chemical potential, which is the conjugate of the central charge, is given by Eq. \eqref{p}
\begin{equation}
\mu=-\frac{S^2-\pi S C+\pi^2 \tilde{Q}^2 - 3 \alpha C^2 \pi^2l^{-2}}{4 \pi^{3 / 2} l C\sqrt{S C}}.
\end{equation}
Finally,  the expression of the conjugate quantity of $\alpha$ or the GB potential is given by
\begin{equation}
	\label{A}
\mathcal{A}= \frac{C^2\, \sqrt{\pi}}{2\, l^3\, \sqrt{C\, S}}.
\end{equation}
We notice that in the RPST formalism, Schwarzschild black holes i.e. $ \tilde{Q}=0$ are equivalent to RN black holes with an effective charge due to the 4D-EGB coupling. However, The Hawking-Page phase transition in Schwarzschild black holes may occurs at a entropy and temperature specific. Electrical and chemical potentials depend on the electric, the central charges, while the GB potential $\mathcal{A}$  depends only on the central charge. Furthermore, the expression of the Gauss-Bonnet action $I_{GB}$ in term of thermodynamics and AdS/CFT quantities by using  Eqs. (\ref{GB}), \eqref{16} and \eqref{A} is given by
\begin{equation}
    I_{GB} =  - \frac{ \alpha \pi^2 C^2 S}{3S^2 l^2 + \pi C S l^2 - \pi^2 \tilde{Q}^2 l^2 - \alpha C^2 \pi^2}.
\end{equation}

This action observes a divergence for $T=0$ as shown in Eq. (\ref{16}). Such unphysical divergences were already found in the on-shell action and surface terms of the total action of the 4D Gauss-Bonnet theory \cite{2020note}.
We can check easily that  the mass of the 4D-EGB black holes is a first-order homogeneity and  $T$, $ \Phi$, $\mu$ and $\alpha$ are zeroth order homogeneity in $S$, $C$, $\mu$ and $\mathcal{A}$. Indeed, a re-scaling $S$, $\tilde{Q}$, C and $\mathcal{A}$ as $S \rightarrow \lambda S$, $ \tilde{Q} \rightarrow \lambda \tilde{Q}$, $C \rightarrow \lambda C$ and $\mathcal{A} \rightarrow \lambda \mathcal{A}$ the mass of the 4D-EGB black hole M, Eq.(\ref{Y}), is re-scaled as  $M \rightarrow\lambda M$ while  $T$, $ \Phi$, $\mu$ and $\alpha$ remain unchanged. This behavior is also observed from the Smarr relation Eq. \eqref{P7}.
\section{Thermodynamic processes}
\label{.4}
This section is devoted to the study of the phase transition and thermodynamic processes  of a charged 4D-EGB black hole by fixing the re-scaled electric charge, the Gauss-Bonnet constant and the central
charge. Critical points of the $T-S$ curve are obtained by inflection points of the temperature  with respect to the entropy at fixed  $\tilde{Q}$, $\alpha$ and C, i.e. by solving the system
\begin{equation}
\label{20}
\left(\frac{\partial T}{\partial S}\right)_{\tilde{Q}, C, \alpha}=0, \qquad \left(\frac{\partial^2 T}{\partial S^2}\right)_{\tilde{Q}, C, \alpha}=0.
\end{equation}
Using Eqs. (\ref{16}) and (\ref{20}), we obtain the critical parameters as
\begin{equation}
    S_c= \frac{\pi C}{6}, \qquad \text{and} \quad \tilde{Q}_c= \pm \frac{C}{l}\, \sqrt{\frac{l^2}{36} - \alpha}.
\end{equation}
Since the charge in the mass expression Eq. \eqref{Y} is raised to the square root, we do not care about the sign of the charge. The critical charge imposes a bound on the parameter i.e  $\alpha l^{-2} < \frac{1}{36}$ and for $\alpha l^{-2} =\frac{1}{36}$, there is no critical charge.  By substituting these critical parameters in  Eqs. (\ref{Y})  and (\ref{16}), we obtain 
\begin{equation}
 M_{c}= \frac{1}{3}\sqrt{\frac{2}{3}} \frac{C}{l} \qquad \text{and} \qquad  T_c = \frac{\sqrt{6}}{3\pi l}.  
\end{equation}
The equation of state (EoS) in the RPST  formalism, by setting  $a= \alpha l^{-2},$ is given by
\begin{equation}
t=\frac{3\,s^2 + 6\, s -q^2 + 36\, a\, \left(  q^2 - 1\right)}{8\, s^{3/2} },
\label{6}
\end{equation}
where we have normalized  the parameters as
\begin{equation}
	t=\frac{T}{T_c}, \quad s=\frac{S}{S_c}, \qquad \text{and} \quad q=\frac{\tilde{Q}}{\tilde{Q}_c}.
\end{equation}
\begin{figure}[htp]
    \centering
    \includegraphics[scale=0.8]{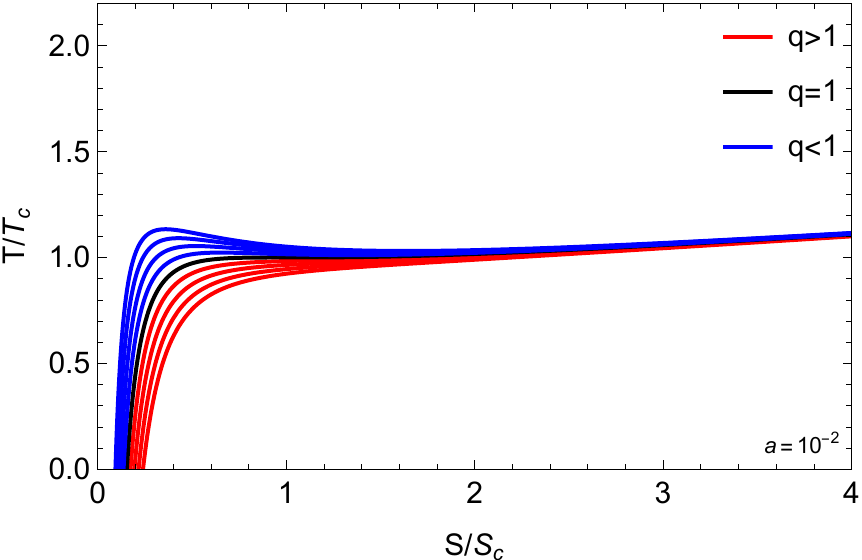}\\
    \includegraphics[scale=0.8]{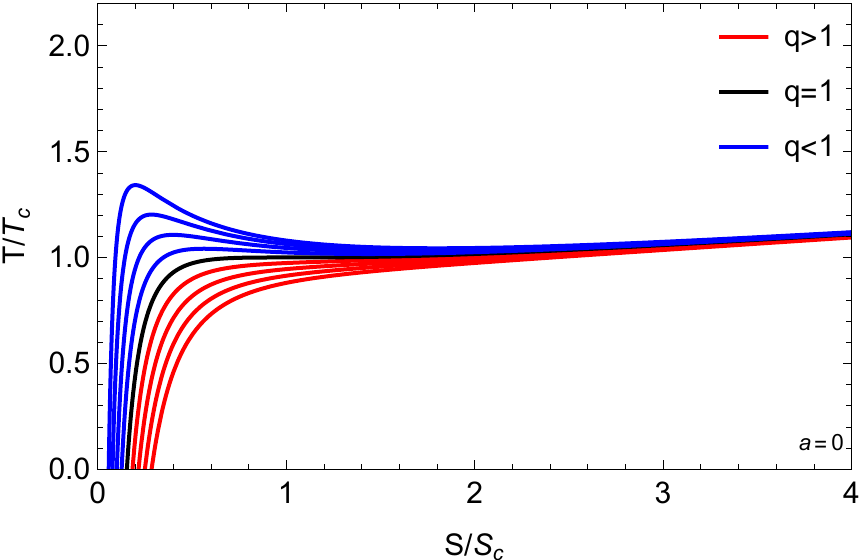}
    \includegraphics[scale=0.8]{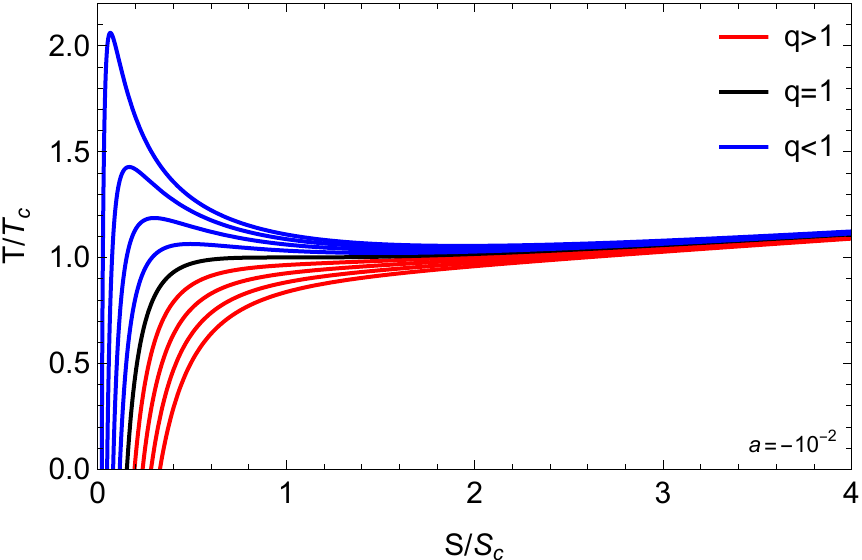}
 \caption{$T-S$ curve for different values of the GB constant and the electric charge.}
    \label{ q=0}
\end{figure}
We can also introduce the Helmholtz free energy F, its the critical parameter and the normalized quantity by
\begin{equation}
F=M - T S,
\end{equation}
\begin{equation}
    F_c= \frac{C}{3\,\sqrt{6}\,l},
\end{equation}
and 
\begin{equation}
    f=\frac{F}{F_c} =\frac{-4 s^{3/2} t+q^2+s^2+6 s}{4 \sqrt{s}}
    \label{8},
\end{equation}
respectively. We notice that, the EoS parameter and the expression of Helmholtz free energy, Eq. (\ref{8}), are  independent from the central charge C. This means that the central charge does not affect the thermodynamic behavior of black holes. This is like "the law of corresponding states" in standard thermodynamics for ordinary matters in which the number has the same behavior.
\\

\begin{figure}[h]
	\centering
	\includegraphics[scale=0.7]{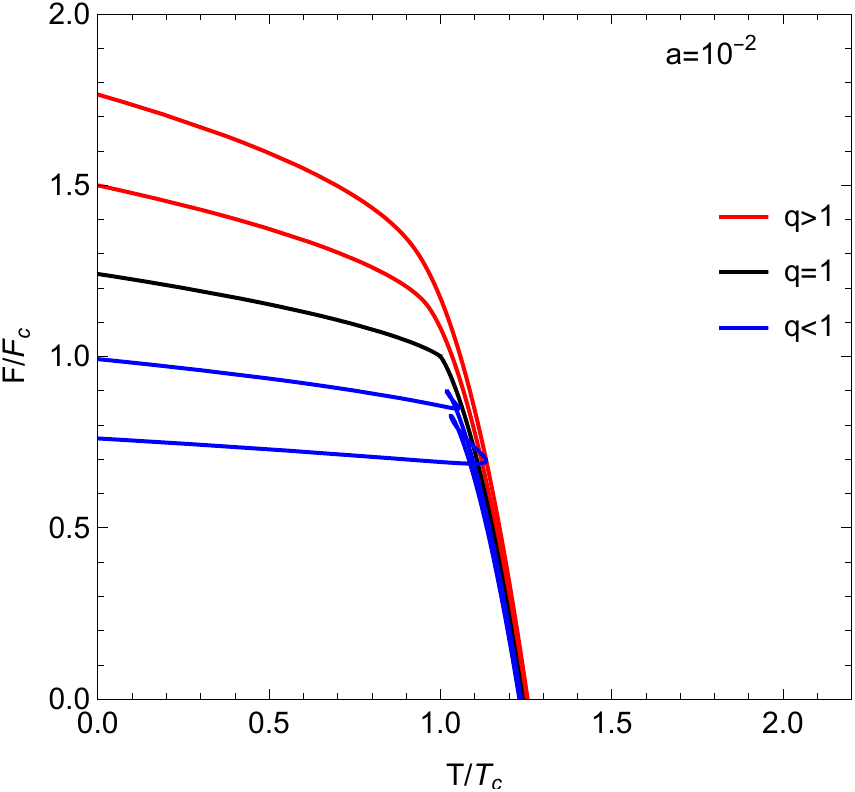}\\
	\includegraphics[scale=0.7]{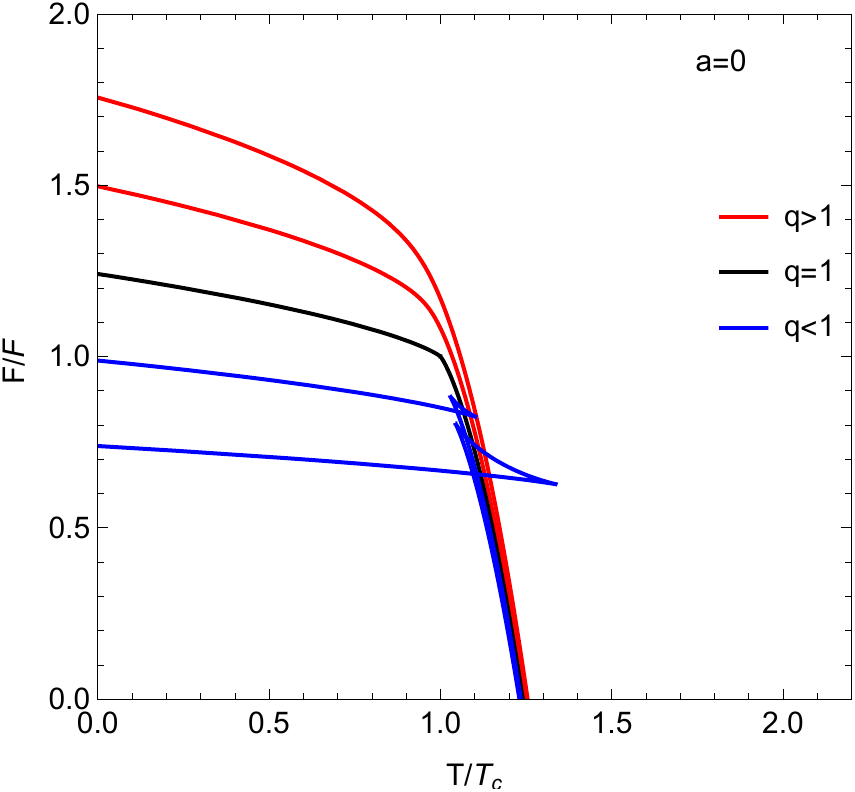}
	\includegraphics[scale=0.7]{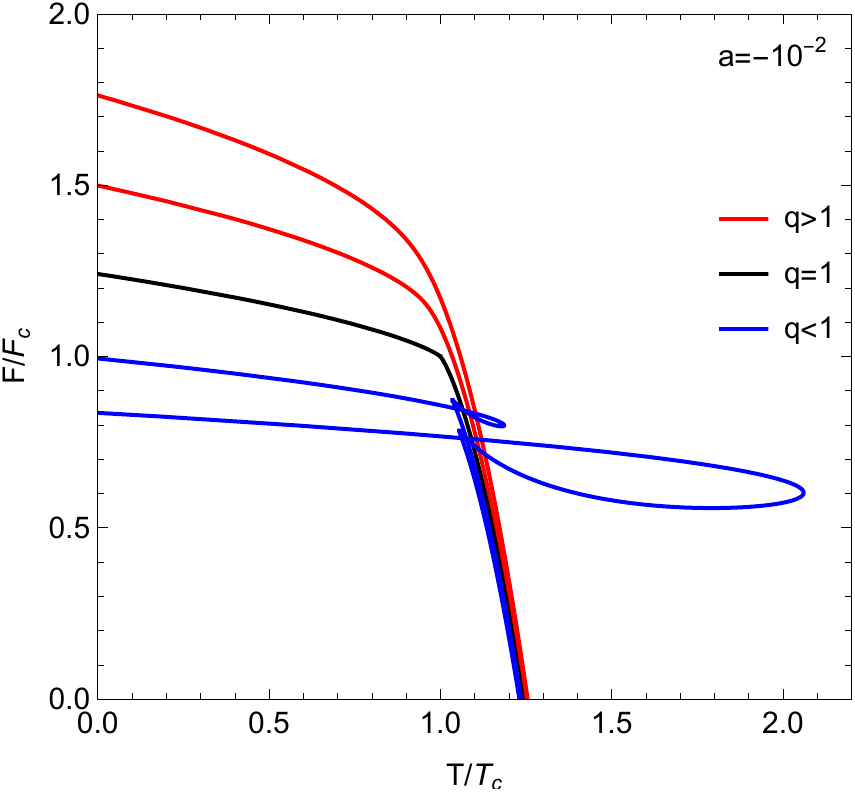}
	\caption{$F-T$ curve for different values of the GB constant and the electric charge. }
	\label{ q=1}
\end{figure}

Figs.\ref{ q=0} and \ref{ q=1} show $T - S$ and $F - T$ curves corresponding to a fixed electric charge and the GB constant  processes for different values of $a$ and $q$. In Fig.1, we notice the same thermal behaviors for $q = 1$,
i.e, $\tilde{Q}= \tilde{Q_c}$. This means that the Gauss-Bonnet coupling constant $\alpha$ has no effect on the thermal evolution of the 4D-EGB black holes and this case is accompanied with a second-order phase transition. This result can also be seen from Eq. (\ref{6}), where RN-AdS black holes at the critical electric charge  are recovered even for 4D-EGB black holes \cite{bGao_2022}. The same is true for the free energy $F-T$ curves in Fig.\ref{ q=1}. For $0<\tilde{Q}<\tilde{Q_c} $,  the $T-S$ curves are non-monotonic.  This represents a first-order phase transition from  small black holes to a large one passing through the medium sized black holes. While the latter is unstable as it corresponds to a transition zone,  small  and  large black holes are stable. The $F-T$ curves, in this case ($0<\tilde{Q}<\tilde{Q_c} $), contain a swallow tail with the root located at  temperature $T > T_c$ (supercritical temperature). The Gauss-Bonnet constant affects the evolution of the $T-S$ and the $F-S$ curves for $0<\tilde{Q}<\tilde{Q_c}$. Indeed, with the same value of $q$, transition temperatures $T_{tran}$,  are different for different negative values of $\alpha$. The transition temperature is, furthermore, more pronounced for a  negative value of $\alpha$  compared to a vanishing or a positive value.  For $\tilde{Q} >\tilde{Q}_c$ there are no phase transitions. 

\section{Uncharged 4D-EGB Black Holes}
\label{.5}
From Eqs. (\ref{16}) and (\ref{6}), we notice that for an uncharged 4D-EGB black hole, the GB constant behaves as an effective charge in the RPST formalism. This means that an uncharged 4D-EGB black holes behaves like a RN-AdS black holes in AdS/CFT.
In this section, we will study an uncharged 4D-EGB black holes in restricted phase space  thermodynamics. The first law and the Smarr relation of an  uncharged 4D-EGB black holes write as 
\begin{equation}
d M=T d S+\mu d C+\mathcal{A} d \alpha, 
\end{equation}
\begin{equation}
M=T S+\mu C+\mathcal{A} \alpha,
\end{equation}
respectively. The expression of the mass of the uncharged 4D-EGB black hole becomes from Eq. \eqref{Y}
\begin{equation}
	\label{47}
M=\frac{S^2+\pi S C + \pi^2 \alpha C^2 l^{-2}}{2 \pi^{3 / 2} l \sqrt{S C}}.
\end{equation}
The conjugate thermodynamics variables are as follows 
\begin{equation}
	\label{48}
T=\frac{3 S^2+\pi S C- \alpha C^2 \pi^2 l^{-2}}{4 \pi^{3 / 2} l S \sqrt{S C}},
\end{equation}
\begin{equation}
\mu=-\frac{S^2-\pi S C-3 \alpha C^2 \pi^2 l^{-2}}{4 \pi^{3 / 2} l C \sqrt{S C}},
\end{equation}
\begin{equation}
\mathcal{A}=\frac{C^2 \sqrt{\pi}}{2 l^3 \sqrt{C S}}.
\end{equation}
Given  the conjugate thermodynamics variable in the RPST formalism, we proceed by studying the phase transition and thermodynamic processes ($T-S$ processes) of an uncharged 4D-EGB black holes with a fixed  Gauss-Bonnet constant and a central charge.
The critical points in the $T-S$ curve at fixed  $\alpha $ and C, are

\begin{equation}
S_c=\frac{\pi C}{6}, \quad \alpha_c=\frac{l^2}{36}, \quad T_c=\frac{\sqrt{6}}{3 \pi l} , \quad \text{and} \quad F_c=\frac{\sqrt{6}\, C}{18 \,l}.
\end{equation}
The equation of state and the Helmhotz free energy in RPST formalism for the uncharged 4D-EGB black holes are
\begin{equation}
t=\frac{3 s^2+6 s-b}{8 s^{3 / 2}},
\end{equation}
and
\begin{equation}
f=\frac{b+s^2+6 s-4 t\, s^{3 / 2}}{4 s^{1 / 2}},
\end{equation}
where we have defined the relative parameter b as $\frac{\alpha}{\alpha_c}$.
\\
\begin{figure}[htp]
\centering
    \includegraphics[scale=0.6]{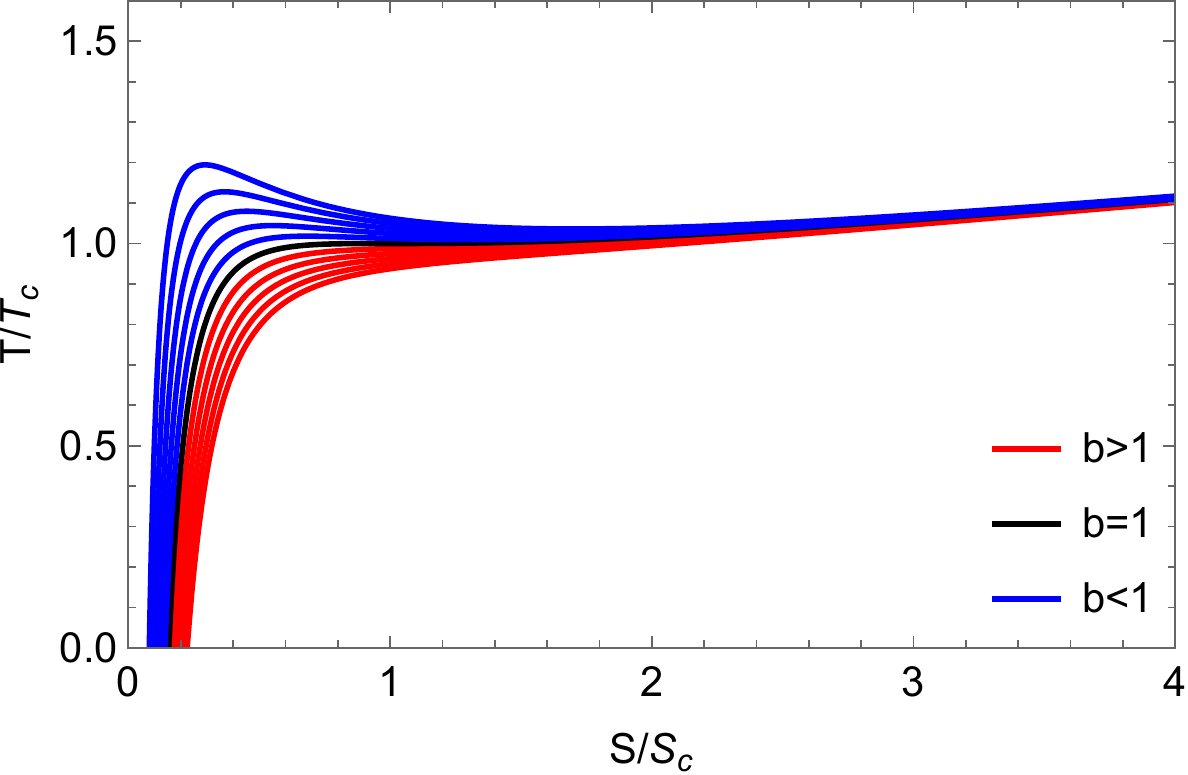}
    \caption{$T-S$ curve for different values of the GB constant.}
    \label{ q=7}
\end{figure}%
\begin{figure}[htp]
	\centering
la	\includegraphics[scale=0.5]{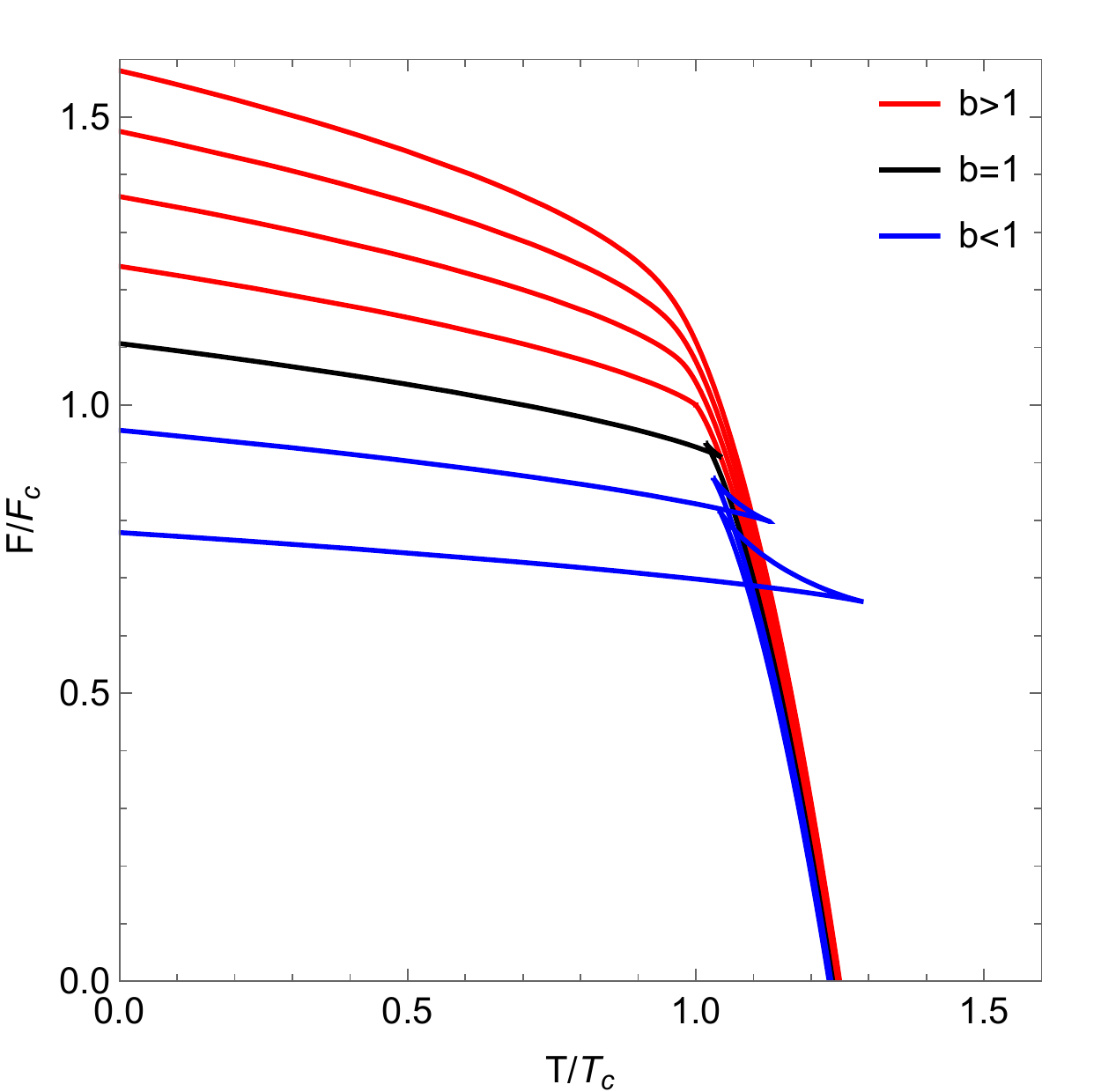}
\caption{$F-T$ curve for different values of the GB constant.}
 \label{ q=5}
\end{figure}
\\
In Figs. \ref{ q=7} and \ref{ q=5}, the $T -S$ and $F- T$ curves correspond to the iso-$\alpha$ processes. Although the black hole is not electrically charged,  it behaves like a RN-AdS  \cite{bGao_2022}. This means that, in the RPST formalism, the Gauss-Bonnet constant causes an uncharged black hole to behave like a charged one. That is, $\alpha$ may be related to an effective electric charge. Indeed, for $\alpha <\alpha_c$,  we have  a first-order phase transition, for $\alpha =\alpha_c$ we have a second-order phase transition, and for $\alpha > \alpha_c$  there is no phase transition.

 To find out this relationship, we consider the mass of RN-AdS black holes  and that of an uncharged 4D-EGB black hole in the RPST formalism. The mass of an uncharged 4D-EGB black hole is
\begin{equation}
M_{4D-EGB}=\frac{S^2+\pi S C + \pi^2 \alpha C^2 l^{-2}}{2 \pi^{3 / 2} l \sqrt{S C}},
\end{equation}
and the one of RN-AdS black holes is 
\begin{equation}
M_{RN-AdS}=\frac{S^2+\pi S C + \pi^2 \tilde{Q}^2}{2 \pi^{3 / 2} l \sqrt{S C}}.
\end{equation}
By comparing the expression of these two masses, a similitude can be made by introducing a re-scaled  effective Gauss-Bonnet charge $\tilde{Q}_{GB}$ as
\begin{equation}
   \tilde{Q}_{GB}= \frac{C\, \sqrt{\alpha}}{l}.
\end{equation}
In the RPST formalism, Schwarzschild black holes in 4D-EGB are similar to those of RN-AdS one.

One of the well known transition between the Schwarzschild black hole and the thermal AdS space is the Hawking-Page phase transition. These transitions may exist with respect to the conservation of charge and angular momentum. In our setup, i.e. the RPST formalism for 4D-EGB gravity, the Schwarzschild black hole becomes charged at thermodynamic properties level. This effective charge has a topological not electromagnetic properties. Thus, we can still consider  the Hawking-Page phase transition between the 4D-EGB black holes and the thermal AdS space in the RPST formalism without violating the conservation of charge and angular momentum. The Gibbs free energy $G$ of the thermal
AdS space is zero. The Hawking-Page phase transition requires  a null Gibbs free energy. 
\begin{equation}
G= M - TS + PV.
\end{equation}
However, as the pressure and the volume are not thermodynamic variables in the RPST formalism, the Gibbs free energy is none other than the Helmholtz free energy F. Putting   G = 0, we get the  Hawking-Page entropy and  temperature from Eqs. \eqref{47} and  \eqref{48}
\begin{equation}
	\label{O8}
 S_{HP}= \frac{\pi  C}{2} \left(\sqrt{\frac{b}{3}+1}+1\right),
\end{equation}
and 
\begin{equation}
	\label{124}
		T_{HP}= \frac{\sqrt{2} \left(\frac{b}{9}+\sqrt{\frac{b}{3}+1}+1\right)}{\pi \, l  \left(\sqrt{\frac{b}{3}+1}+1\right)^{3/2}},    
\end{equation}
respectively. From Figs. \ref{ q=7} and \ref{ q=5} and Eqs. \eqref{O8}-\eqref{124}, we infer that $S_{HP}/S_{c} \leqslant 6$ and $T_{HP}/T_{c}$  is approximately equal to 1.25, i.e, the Hawking-Page phase transition is between a large black hole and thermal  AdS space.
\section{Discussion and Conclusion}
\label{.7}
In this paper, we have studied the 4D-EGB black hole in the context of restricted phase space thermodynamics. We have started by solving the space-time black hole with a negative cosmological constant, with respect to 4D-EGB gravity. From the metric, we deduced the expression of the mass of the black hole in terms of the parameters of the black hole, the geometry, and the GB constant.\\

We constructed a holographic thermodynamic  of the 4D-EGB black hole, based on the AdS/CFT correspondence by equating the partition function of AdS to that of CFT ($Z_{AdS}=Z_{CFT}$) and the free energy $W= \mu C$, to the Euclidean action   $I_E = -\ln{Z_{CFT}}$. We were able to formulate  Smarr relation, the first law and the Gibbs-Duhem relation for a charged and an uncharged 4D-EGB black holes. We show that Smarr relation is homogeneous at a first order. In addition, we have derived thermodynamic variables in the RPST formalism.\\

We have also studied thermodynamic properties of a charged black hole, and we found that the equation of state does not depend on the central charge. This result is similar to that in  the law of corresponding states in standard thermodynamics for ordinary matter. We show that when the charge of black hole is greater than the critical one, there is no phase transition, but when $Q=Q_c $ we find a phase transition of the second order. For $0<Q<Q_c $, we find a first-order phase transition. A remarkable result that not only the charge that influences  thermodynamic properties but even the GB constant, precisely when the charge is sub-critical.\\

For the study of the uncharged black hole, we find that its thermodynamic properties depend on the value of the GB constant. Indeed, for $\alpha = \alpha_c$, it corresponds to a second-order phase transition and   $0<\alpha<\alpha_c $, it corresponds to a first-order phase transition. But, for $\alpha> \alpha_c$, there is no phase transition. Although it is not charged, the black hole in 4D-EGB behaves as if it is  charged  one with an effective charge that depends on the GB constant and the central charge.\\

Finally, the Hawking-Page phase transition may occurs between a large black hole and a thermal AdS space.

\newpage

\bibliography{biblio} 
\bibliographystyle{unsrt}

\end{document}